\documentclass[conference]{IEEEtran}
\IEEEoverridecommandlockouts
\usepackage{cite}
\usepackage{amsmath,amssymb,amsfonts}
\usepackage{algorithmic}
\usepackage{graphicx}
\usepackage{textcomp}
\usepackage{multirow}
\usepackage{tabularx}
\usepackage[table,xcdraw]{xcolor}
\usepackage[normalem]{ulem}
\useunder{\uline}{\ul}{}
\def\BibTeX{{\rm B\kern-.05em{\sc i\kern-.025em b}\kern-.08em
    T\kern-.1667em\lower.7ex\hbox{E}\kern-.125emX}}
\begin{document}

\title{5G Networks Supported by UAVs, RESs, and RISs\\
% {\footnotesize \textsuperscript{*}Note: Sub-titles are not captured in Xplore and should not be used}
}

\author{\IEEEauthorblockN{Adam Samorzewski}
\IEEEauthorblockA{
\textit{Poznan University of Technology} \\
Poznan, Poland \\
adam.samorzewski@doctorate.put.poznan.pl}
\and
\IEEEauthorblockN{Adrian Kliks}
\IEEEauthorblockA{
\textit{Poznan University of Technology}, Poznan, Poland \\ 
and \textit{Luleå University of Technology}, Luleå, Sweden\\
adrian.kliks@put.poznan.pl or adrian.kliks@ltu.se}
}

\maketitle

\begin{abstract}
\label{abstract}
This paper presents the examination of the 5G cellular network aware of Renewable Energy Sources (RESs) and supported by Reconfigurable Intelligent Surfaces (RISs) and Unmanned Aerial Vehicles working as mobile access nodes. The investigations have been focused on the energy side of the Radio Access Network (RAN) placed within the area of the city of Poznan (Poland). The gain related to enabling RES generators, i.e., photovoltaic (PV) panels, for base stations (BSs) was presented in the form of two factors -- the average number of UAV replacements (ANUR) with a fully charged one to ensure continuous access to mobile services for currently served user equipment (UE) terminals, and the average reduction in energy consumption (AREC) within the whole network.\footnote{Copyright © 2023 IEEE. Personal use is permitted. For any other purposes, permission must be obtained from the IEEE by emailing pubs-permissions@ieee.org. This is the author’s version of an article that has been published in the proceedings of the 2023 International Conference on Software, Telecommunications and Computer Networks (SoftCOM) by the IEEE. Changes were made to this version by the publisher before publication, the final version of the record is available at: https://dx.doi.org/10.23919/SoftCOM58365.2023.10271683. To cite the paper use: A. Samorzewski and A. Kliks, "5G Networks Supported by UAVs, RESs, and RISs," \textit{2023 International Conference on Software, Telecommunications and Computer Networks (SoftCOM)}, Split, Croatia, 2023, pp. 1--6, doi: 10.23919/SoftCOM58365.2023.10271683 or visit https://ieeexplore.ieee.org/document/10271683.}
% in the form of counts of UAVs' batteries recharging needed to ensure continuous access to mobile services for all active user equipment terminals (UEs).during the whole simulation time
\end{abstract}


\section{Introduction}
\label{section:introduction}
Nowadays, telecommunication systems are powered mainly by fossil fuels. Those systems contribute about $25\%$ to the total value of carbon dioxide ($\text{CO}_2$) emissions caused by the Information and Communication Technology (ICT) segment, which shows an increasing trend in energy demand from one year to another~\cite{I}. 
% This can be observed for instance comparing the power consumption of $4$G and $5$G base stations (BSs), whereby improving the network link throughput (around $16$ times) the energy utilization has also adequately risen (around $4$ times)~\cite{I}. Although this in fact indicates progress in the energy efficiency of mobile systems achieved throughout the years, there is still a need to provide much more resources to obligatory base hardware. 
Furthermore, currently, the ICT sector is responsible for a huge part of global Green House Gas (GHG) emissions, which are maybe underestimated and could actually be even as high as $2.1$ to $3.9\%$~\cite{Freitag}. Hence, to deal with this issue the necessity of finding alternative sources of power that will both meet the energy demand of wireless systems ($4$G, $5$G, and beyond) and reduce the amount of produced $\text{CO}_2$ emissions to the atmosphere might be required. Therefore, the engagement of Renewable Energy Sources (RESs), e.g., photovoltaic (PV) panels seems to be an appropriate solution~\cite{Deruyck}. However, it is still hard to be unequivocally stated whether the use of solar cells is much more environment-friendly (in terms of entailing emissions) or not, especially taking into account the processes of their production and utilization. On the contrary, the way of harvesting electrical energy from solar radiation seems to be non-polluting itself and the amount of potentially generated resources can be considered to some extent as endless (excluding the need to replace/renovate PV panels after some time period) in a long-term context~\cite{Abid}. 
% Nevertheless, due to the impact of weather conditions on the intensity of getting energy resources by solar panels at a specific time moment, it could be necessary to implement a control system for cellular networks to ensure sustainable (radio and energy) resource management, e.g., in the form of operating algorithms of traffic steering, resource allocation, etc.

Within the concept of cellular systems of the $5^\text{th}$ generation, mobile services are divided into $4$ main groups: Enhanced Mobile Broadband (eMBB), Critical Communications (CC) and Ultra Reliable Low Latency Communications (URLLC), Massive Internet of Things (mIoT), and flexible network operations, where first $3$ of them, in short, assume the provision of high throughput, low latency, and a big number of connected devices at the same time, respectively~\cite{3GPP:TR21.915}. From the perspective of delivering the first case, there may appear a limitation related to a finite number of physical resource blocks (PRBs) per network cell. This issue might be noticed especially in urban areas, where the population is quite dense. Thus, to guarantee sufficient capacity and bit rate in a given area, the idea of deploying base stations covering regions with so-called small cells has been developed. This idea of wireless system implementation is based on the provision of RAN with low-power access nodes distributed very close to each other. However, while bringing this small cell concept to life due to obstacles such as unfavorable city architecture or money deficits for building new stationary base stations, there is a risk of signal gap appearance. One of the approaches to handle this issue, which is more and more often taken under consideration in scientific works is the employment of additional supportive equipment, e.g., Unmanned Aerial Vehicles (UAVs) as mobile base stations (MBSs). This, in turn, gives mobile network operators (MNOs) the ability to dynamically adjust the actual locations of access nodes to cover radio signal gaps and/or support existing telecommunication infrastructure in serving areas (urban or remote), where the number of simultaneously connected user terminals can also fluctuate and even exceed primary assumed capacity (e.g., due to public events)~\cite{Mozaffari, Alzenad}.

Besides, the use of so-called Reconfigurable Intelligent Surfaces (RISs) to control the radio signal coverage of wireless systems has got great attention in the current literature. The RIS is a device in the form of a surface consisting of a large number of passive (or sometimes active) reflecting elements, which are able to independently cause a desired change in phase and/or amplitude of the incident radio signal. As a consequence of assuring flexible reconfiguration of signal propagation (after enabling RISs), MNOs would be able to accomplish better performance of their networks by reducing interferences and dropouts and raising the reliability, capacity, and throughput of radio links~\cite{Huang, Di Renzo}.

Thus, in this paper, we evaluate the true performance of the UAVs working as base stations, which are equipped with both RISs (for potential future use) and RESs (for extension of the operating time). The goal of this analysis is to find the most appropriate deployment and setup of the drone base stations over a certain area while considering the transmission power, the energy consumed by UAVs for their operation, and the energy production by the RESs. We evaluate the trade-off between the additional mass needed to carry RES and RIS elements and the gains related to energy generated by RESs. Finally, we carry out simulations based on real atmospheric data to verify the true performance of such a solution. 

\section{Scenario}
\label{section:scenario}
\begin{figure}[!h]
\centering
\includegraphics[width=0.475\textwidth]{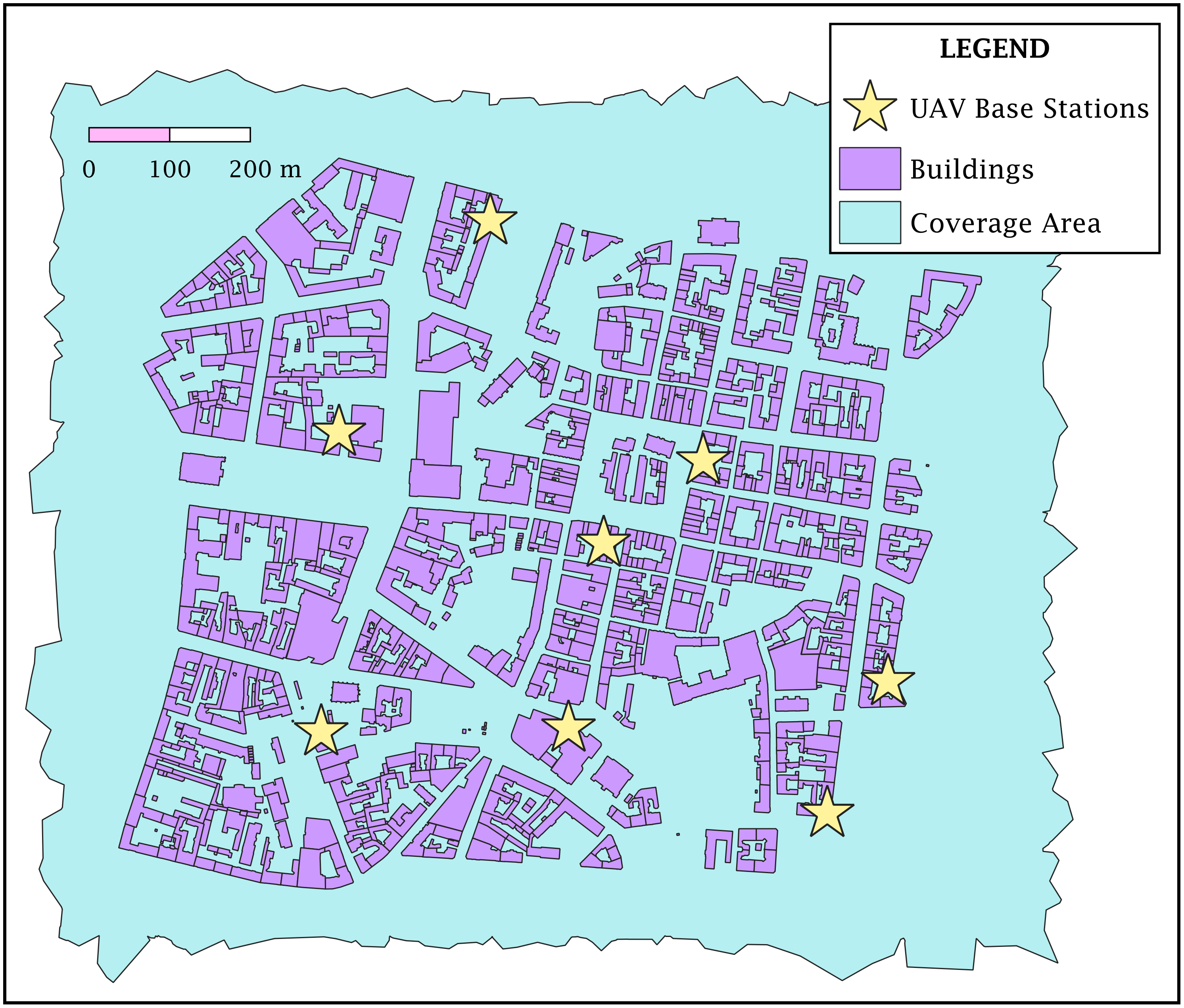}
\caption{Map of the examined area within the city of Poznan.}
\label{figure:map}
\end{figure}

The scenario considered in the work takes into account the $5$G cellular network placed within the old market of the city of Poznan in Poland (data given in \cite{PoznanData}), the base stations of which are UAVs located based on the real data of one of the Polish mobile operator (given in \cite{NetworkData}) and hovering $50$ m above the ground. For this study, there was assumed that the placements of the drones do not change over time. Around the city, there are $100$ outdoor users distributed randomly with fixed bit rate requirements equal to $100$ Mbps per each. In Fig.~\ref{figure:map} the map presenting the above-described scenario of the wireless system has been attached.
%475

\subsection{Network Design}
\label{subsection:design}
The examined network has been designed according to the planner tool described in \cite{Castellanos} named Green Radio Access Network Design (GRAND). This tool projects and optimizes radio access networks toward power consumption and/or human exposure by enabling the optimal number of cells of base stations with predefined locations and adjusting their transmission parameters based on the instantaneous throughput requirements of user terminals connected to the wireless system. As the input data, the GRAND tool receives the lists of available base stations and active users with their specific transmission data, and the shape files describing the considered environment (in $3$D) including buildings and coverage area. The location of each user is chosen randomly within the provided simulator for every single run. In addition, in the very beginning, all transceivers of access nodes broadcast the radio signal with maximum power. After distributing users within the zone of study, the planner software selects which one of the network cells shall be enabled and with what power they shall transfer the data to reach as many active users as possible. Within this step, the GRAND tool assesses potential radio links between the system and each UE based on the bit rate requested by it as well as the maximum allowable path loss for a particular association for different transmit power configurations. Next, when the graph of accessible BS-UE relations is prepared, a mixed integer programming (MIP) solver is engaged to search for the optimal result. In order to obtain this result an objective function is formulated based on the envisioned optimization of human exposure, power consumption, or both. Afterward, all the users start exchanging data with appropriate access nodes in a continuous manner causing a fixed traffic load throughout the whole time of a single simulation run. Simultaneously, energy balance calculations for all base stations are done, which are further translated to an average number of needed battery recharging per UAV with and without RESs.

\subsection{Equipment}
\label{subsection:equipment}
The implementation of PV panels within the GRAND tool software has been inspired by the specifications of the real device, which can be found in \cite{PvData}. However, it has been assumed that used PV panels are mounted on the top of the UAV's cover in the form of thin-film solar cells. Thus, because of the insignificant impact on the total power consumption of a single mobile base station, the weight of PV panels was omitted during all calculations devoted to energy characteristics within a simulation run. In addition, those calculations have been performed for $4$ various days (each starting a different season of the year) to indicate how the harvesting of energy by PV panels depends on the time of the year in Poland.

Similarly, as for solar panels, the battery systems for UAV BSs have been also designed based on the real implementation described in \cite{BatteryData}. The objectives of using batteries for drones are to power them in order to fulfill the energy demand as well as to store the resources produced by RES generators connected to them. The lack of electrical energy delivered by the battery cells entails a need to replace the UAV with another one but fully charged. Nevertheless, it has been posited that batteries can be completely drained, and 5\% of the total energy kept in a single one is always used for flights before and after the service.

The transceivers of UAV access nodes work in accordance with the Multiple-Input-Multiple-Output (MIMO) technology using $64$ active antenna elements (AAE) transmitting radio signals in the $3500$ MHz frequency band. The methods of channel estimation and signal processing adopted in the software are consistent with the minimum mean-squared error (MMSE) schema. Furthermore, each of those mobile base stations has also one device of RIS type with $16$ identical passive reflecting elements performing phase shifting with a $6$-bit resolution of the impinging radio signal. For the following study, the impact of RISs implementation within the examined area on radio signal propagation has been neglected contrary to its influence on the energy balance of the wireless system. However, their effect on the performance of the network will be taken into consideration in future work.

\subsection{Energy Models}
\label{subsection:models}
All energy models used to provide the mathematical calculations within the GRAND tool (related to energy production and consumption -- prosumption) have been directly taken from or inspired by scientific literature or real implementations. 
% The list of those formulas has been contained in Tab.~\ref{table:models}. 

% \begin{table}[!h]
% \centering
% \label{table:models}
% \caption{The list of energy models implemented within the GRAND tool}
% \resizebox{0.48\textwidth}{!}{
% \begin{tabular}{|l|c|c|}
% \hline
% \multicolumn{1}{|c|}{Name of Energy Model}              & Sign            & Source  \\ \hline
% Power Consumption of Multirotor UAV (hovering)          & $P_\text{UAV}$  & \cite{} \\ \hline
% Power Consumption of MIMO Transceiver                   & $P_\text{MIMO}$ & \cite{} \\ \hline
% Power Consumption of RIS Array                          & $P_\text{RIS}$  & \cite{} \\ \hline
% Output Power of PV Panels                               & $P_\text{PV}$   & \cite{} \\ \hline
% Gravitational Acceleration at Altitude                  & $g$             & \cite{} \\ \hline
% Air Density at Altitude                                 & $\rho$          & \cite{} \\ \hline
% \end{tabular}}
% \end{table}

\subsubsection{UAV Device}
\label{subsubsection:uav}
Due to the system design concept, it was assumed that UAV access nodes do not change their positions during the simulation. Thus, within the evaluation of energy consumption related to the movement, there is only a need to assess the utilization caused by UAV hovering $\left(P_\text{UAV}\right)$. Thus, according to \cite{Janji}, we present the following formula: 
\begin{align}
    \label{equation:uav}
    P_\text{UAV}\left(t\right)=\sqrt{\frac{\Big(\big(m_\text{UAV}+m_\text{PKG}\big)\cdot g\left(t, h_\text{UAV}\right)\Big)^{3}}{2\cdot\pi \cdot r_\text{p}^{2}\cdot l_\text{p}\cdot\rho\left(t, h_\text{UAV}\right)}},
\end{align}
where $t$ is the current time step and $m_\text{UAV}$ and $m_\text{PKG}$ are the masses of the UAV and the package that is lifted by it. In our case $m_\text{PKG}=N_\text{MIMO}\cdot m_\text{MIMO}+N_\text{RIS}\cdot m_\text{RIS}+N_\text{PV}\cdot m_\text{PV}+m_\text{AUX}$, where $m_\text{MIMO}$, $m_\text{RIS}$, and $m_\text{PV}$ are the masses of a single transceiver, RIS array, and PV panel, and $N_\text{MIMO}$, $N_\text{RIS}$, and $N_\text{PV}$ are their numbers, respectively. The parameter of $m_\text{AUX}$ is the mass of the auxiliary hardware/package carried by the UAV. Next, $l_\text{p}$ and $r_\text{p}$ are the number of UAV propellers and the radius of a single one, adequately. Finally, $g$ and $\rho$ are the gravitational acceleration and air density at the altitude $h_\text{UAV}$ and time moment $t$.

\subsubsection{RIS Array}
\label{subsubsection:ris}
Each UAV base station is equipped with a single RIS array in order to improve the efficiency of radio signal broadcasting by its reflecting. Although within this study the use of RISs is omitted for the propagation (human exposure) side of the network design, as a form of preparation for further investigations, the examination of the impact of employing RISs on the energy demand of the wireless system $\left(P_\text{RIS}\right)$ has been taken into considerations as well. Hence, referring to \cite{Huang}, we propose the model attached below:
\begin{align}
    \label{equation:ris}
    P_\text{RIS}\left(t\right)=N_\text{RIS}\cdot N_\text{RE}\cdot P_\text{PSH}\left(b_\text{PSH}\right),
\end{align}
where $N_\text{RIS}, $$N_\text{RE}$ are the numbers of used arrays and identical reflecting elements per single RIS, which effectively perform phase shifting on the impinging signal. Next, $P_\text{PSH}$ is the power consumption of each phase shifter, which is dependent on the bit resolution $b_\text{PSH}$ of the used type.

\subsubsection{MIMO Transceiver}
\label{subsubsection:mimo}
The model used to estimate the power consumed by radio hardware has been formulated in accordance with the work contained in \cite{Björnson}. The mathematical formula that evaluates the total power consumption $\left(P_\text{MIMO}\right)$ by a single transceiver in the current time step $t$ is as below:
\begin{align}
    \label{equation:power_mimo}
    P_\text{MIMO}\left(t\right)=P_\text{CP}\left(t\right)+P_\text{PA}\left(t\right),
\end{align}
where $P_\text{PA}\left(t\right)=\frac{P_\text{TX}\left(t\right)}{\mu_\text{PA}}$ is the power consumed by the power amplifier. The parameters of $P_\text{TX}$ and $\mu_\text{PA}$ are the transmit power and efficiency of the amplifier, respectively. Next, the $P_\text{CP}$ is the power spent by circuit components of the transceiver, which is expressed as follows: 
% The latter refers to additional equipment not related to providing the radio connection (e.g., lighting). The former was expressed as follows:
\begin{align}
    \label{equation:power_circuit}
    \nonumber
    % \frac{P_\text{TX}}{\mu_\text{PA}}+
    &P_\text{CP}\left(t\right)=P_\text{FIX}+P_\text{TC}\left(t\right)+P_\text{CE}\left(t\right)+P_\text{C/D}\left(t\right)+P_\text{BH}\left(t\right)\\&+P_\text{SP}\left(t\right),
\end{align}
where $P_\text{FIX}$ is the fixed power consumed by a cell node. The parameter of $P_\text{TC}\left(t\right)=M_\text{BS}\left(t\right) P_\text{CC}$ is the power utilized by the transceiver chains in the time step $t$, where $P_\text{CC}$ is the power that is required to run the circuit components (e.g. filters, I/Q mixers, etc.), and $M_\text{BS}$ is the number of presently active antenna elements of the cell. Next, $P_\text{CE}\left(t\right)=\frac{3B_\text{w}}{\tau_\text{c}\cdot \eta_\text{BS}}K_\text{UE}\left(t\right)\left(M_\text{BS}\left(t\right)\tau_\text{p}\left(t\right)+M_\text{BS}^{2}\left(t\right)\right)$ is the power needed by the channel estimators, which work according to the minimum mean-squared error (MMSE) scheme \cite{Björnson}. The parameters of $B_\text{w}$ and $\eta_\text{BS}$ are the channel bandwidth and computational efficiency, respectively. In addition, $\tau_\text{p}\left(t\right)=\text{RF}\cdot K_\text{UE}\left(t\right)$ is the number of samples allocated for pilots per coherence block in a specific time step $t$, where $\text{RF}$ is the pilot reuse factor and $K_\text{UE}$ is the current number of served users. In turn, $\tau_\text{c}=B_\text{c}\cdot t_\text{c}$ is the number of samples per coherence block, where $B_\text{c}$ and $t_\text{c}$ are the coherence bandwidth and time, respectively. Furthermore, 
% $P_\text{C/D}\left(\text{TR}_\text{UL}, \text{TR}_\text{DL}\right)=P_\text{COD}\cdot\text{TR}_\text{DL}+P_\text{DEC}\cdot\text{TR}_\text{UL}$
$P_\text{C/D}\left(t\right)=P_\text{COD}\cdot\text{TR}_\text{DL}\left(t\right)+P_\text{DEC}\cdot\text{TR}_\text{UL}\left(t\right)$
is the total power consumed by a transceiver of the UAV BS for encoding $\left(P_\text{COD}\right)$ and decoding $\left(P_\text{DEC}\right)$ the information transferred through uplink, in short UL, $\left(\text{TR}_\text{UL}\right)$ and downlink, in short DL, $\left(\text{TR}_\text{DL}\right)$ connections in a particular time step $t$. The power model takes into account also the load-aware part of the consumption referring to the backhaul links -- $P_\text{BH}\left(t\right)=P_\text{BT}\big(\text{TR}_\text{UL}\left(t\right)+\text{TR}_\text{DL}\left(t\right)\big)$, where $P_\text{BT}$ is the backhaul traffic power. Finally, the power required by the network cell for operations related to signal processing (e.g., UL reception and DL transmission, computation of the combining/precoding vectors) compliant with the MMSE scheme $\left(P_\text{SP}\right)$ is as below: 
% by the following equation: 
\begin{align}
    \label{equation:power_signal_processing}
    % \frac{P_\text{TX}}{\mu_\text{PA}}+
    &P_\text{SP}\left(t\right)=\frac{3B_\text{w}}{\tau_\text{c}\cdot \eta_\text{BS}}\Bigg[M_\text{BS}\left(t\right) K_\text{UE}\left(t\right) \big(\tau_\text{u}\left(t\right)+\tau_\text{d}\left(t\right)\big)\\ \nonumber &+\frac{\left(3M_\text{BS}\left(t\right)^{2}+M_\text{BS}\left(t\right)\right)K_\text{UE}\left(t\right)}{2}+\frac{M_\text{BS}\left(t\right)^{3}}{3}+2M_\text{BS}\left(t\right)\\ \nonumber &+M_\text{BS}\left(t\right)\tau_\text{p}\left(t\right)\big(\tau_\text{p}\left(t\right)-K_\text{UE}\left(t\right)\big)+M_\text{BS}\left(t\right) K_\text{UE}\left(t\right)\Bigg],
\end{align}
where $\tau_\text{d}\left(t\right)$ is the number of DL data samples per coherence block in the time step $t$. 

\subsubsection{PV Panel}
\label{subsubsection:pv}
There is also a need to model the energy harvesting process performed by PV panels mounted on the cover of each UAV access node. The output power of the set of PV arrays $\left(P_\text{PV}\right)$ in a certain time step $t$ is denoted as \cite{HomerPro:v3.15}: 
\begin{align}
    \label{equation:power_pv}
    % \nonumber
    P_\text{PV}\left(t\right)=N_\text{PV} P_\text{R,PV} f_\text{PV}\cdot\frac{\overline{G}_\text{T}\left(t\right)}{\overline{G}_\text{T,STC}}\bigg[1+\alpha_\text{P}\Big(T_\text{c}\left(t\right)-T_\text{c,STC}\Big)\bigg],
\end{align}
where $N_\text{PV}$, $P_\text{R,PV}$, and $f_\text{PV}$ are the total number of PV panels allocated per network cell, and the rated power and derating factor of a single one. In addition, the first one is the multiplication of the numbers of PV panels connected in series $\left(N_\text{PV,s}\right)$ and parallel $\left(N_\text{PV,p}\right)$. 
% The latter, in turn, is specified by $f_\text{PV}=(\varphi_\text{DCC})^{\psi}(\varphi_\text{IT}\cdot\varphi_\text{AC})^{\psi-1}\varphi_\text{DC}\cdot\varphi_\text{MT}\cdot\varphi_\text{MM}\cdot\varphi_\text{CON}\cdot\varphi_\text{SO}\cdot\varphi_\text{SD}\cdot\varphi_\text{SH}$, where the parameters marked with $\varphi$ are the factors related to power losses of the DC converter, AC-DC inverter/transformer, AC and DC wiring, PV module tolerance and mismatch, diode connections, system downtime, soling, and shading, respectively. The parameter of $\psi$ is the flag signaling the PV system delivers DC $(\psi=1)$ or AC $(\psi=0)$ current. 
Next, $\overline{G}_\text{T}$ and $T_\text{c}$ are the parameters that denote the solar radiation incident on the PV array and its temperature. Thus, $\overline{G}_\text{T,STC}$ and $T_\text{c,STC}$ define the values of the same parameters but for standard test conditions (STC). Finally, $\alpha_\text{P}$ is the temperature coefficient of power dependent on the type of used PV panels. Besides, to assess the temperature of the PV cell $\left(T_\text{c}\right)$, the formula is used \cite{HomerPro:v3.15}:
\begin{align}
    \label{equation:temperature_solar_cell}
    % \nonumber
    &T_\text{c}\left(t\right)=\frac{T_\text{a}\left(t, h_\text{PV}\right)}{1+\left(T_\text{c,NOCT}-T_\text{a,NOCT}\right)\left(\frac{\overline{G}_\text{T}\left(t\right)}{\overline{G}_\text{T,NOCT}}\right)\left(\frac{\alpha_\text{P}\mu_\text{mp,STC}}{\tau\alpha}\right)} \\ \nonumber 
    +&\frac{\left(T_\text{c,NOCT}-T_\text{a,NOCT}\right)\left(\frac{\overline{G}_\text{T}\left(t\right)}{\overline{G}_\text{T,NOCT}}\right)\left[1-\frac{\mu_\text{mp,STC}\left(1-\alpha_\text{P}T_\text{c,STC}\right)}{\tau\alpha}\right]}{1+\left(T_\text{c,NOCT}-T_\text{a,NOCT}\right)\left(\frac{\overline{G}_\text{T}\left(t\right)}{\overline{G}_\text{T,NOCT}}\right)\left(\frac{\alpha_\text{P}\mu_\text{mp,STC}}{\tau\alpha}\right)},
\end{align}
where $T_\text{c,NOCT}$, $T_\text{a,NOCT}$, and $\overline{G}_\text{T,NOCT}$ are the nominal operating cell temperature (NOCT) of the PV panel, and the ambient temperature and solar radiation at which the NOCT is defined, respectively. Next, $\tau$, $\alpha$, and $\mu_\text{mp,STC}$ are the solar transmittance of any cover over the PV array and its solar absorptance, and the maximum power point efficiency of the PV panel under STC. This efficiency is equal to $\mu_\text{mp,STC}=\frac{P_\text{R,PV}}{a_\text{PV}\cdot b_\text{PV}\cdot\overline{G}_\text{T,STC}}$, where $a_\text{PV}$ and $b_\text{PV}$ are the dimensions of a single PV module. In turn, the parameter of $h_\text{PV}$ is the ground-relative altitude of the PV panels powering a specific cell.

\subsubsection{Battery System}
\label{subsubsection:battery}
In order to shape the energy management within the battery system $\left(E_\text{BATT}\right)$ of each UAV, inspired by \cite{Voltacon} we propose a new model specified below:
\begin{equation}
    \label{equation:battery_state}
    E_\text{BATT}\left(t\right)=
    \begin{cases}
        E_\text{BATT}\left(t'\right)+\Delta E_{\text{BATT},1}\left(t\right), & \text{if } \Delta E\left(t\right) > 0\\
        E_\text{BATT}\left(t'\right)+\Delta E_{\text{BATT},2}\left(t\right), & \text{otherwise}
    \end{cases},
\end{equation}
where $E_\text{BATT}\left(t'\right)$ is the energy handled by the battery system in the previous time step $t'$. Next, $\Delta E_{\text{BATT},1}$ and $\Delta E_{\text{BATT},2}$ are the energy amounts that have to be transferred from/to the battery system in the current time step $t$. Finally, $\Delta E$ is the energy balance, i.e., the difference between required energy and harvested one at the same moment. The parameters of $\Delta E_{\text{BATT},1}$ and $\Delta E_{\text{BATT},2}$ can be expressed by the formulas:
\begin{align}
    \label{equation:battery_delta_1}
    &\Delta E_{\text{BATT},1}\left(t\right)\\ \nonumber&=\text{max}\Big(\Delta E\left(t\right)\cdot\mu_\text{BATT},E_\text{BATT,max}-E_\text{BATT}\left(t'\right)\Big), 
\end{align}
\begin{align}
    \label{equation:battery_delta_2}
    \Delta E_{\text{BATT},2}\left(t\right)=\text{max}\left(\frac{\Delta E\left(t\right)}{\mu_\text{BATT}},-E_\text{BATT}\left(t'\right)\right),
\end{align}
where $\mu_\text{BATT}$ and $E_\text{BATT,max}$ are the efficiency of the used battery type and maximum energy the battery system is able to collect, respectively. The latter is equal to $E_\text{BATT,max}=N_\text{BATT}E'_\text{BATT,max}$, where $E'_\text{BATT,max}$ is the maximum energy of a single battery, and $N_\text{BATT}=N_\text{BATT,s}N_\text{BATT,p}$ is the total number of accumulator units in a battery system. The parameters of $N_\text{BATT,s}$ and $N_\text{BATT,p}$ are the numbers of batteries linked to each other in serial and parallel order, respectively. To evaluate the current energy balance $\left(\Delta E\right)$, the formula below was engaged:
\begin{align}
    \label{equation:battery_delta}
    &\Delta E\left(t\right)=\\ \nonumber&\Bigg(P_\text{PV}\left(t\right)-\frac{P_\text{UAV}\left(t\right)+P_\text{MIMO}\left(t\right)+P_\text{RIS}\left(t\right)}{1-\sigma_\text{DC}}\Bigg)\Bigg(t-t'\Bigg),
\end{align}
where $\sigma_\text{DC}$ is the loss factor related to DC supplying the hardware parts of the UAV device.

\subsubsection{Atmospheric Parameters}
\label{subsubsection:atmosphere}
Finally, let us collect all auxiliary formulas used to calculate necessary atmospheric parameters. The air density $\left(\rho\right)$ at the altitude $h$ and in the current time step $t$ can be calculated as follows \cite{Omni}:
\begin{align}
    \label{eqAirDensity}
    &\rho\left(t, h\right)=\\ \nonumber &\frac{p_\text{d}\left(t, h\right)}{R_\text{d}\cdot \big(T_\text{a}\left(t, h\right)+273.15\big)}+\frac{p_\text{v}\left(t, h\right)}{R_\text{v}\cdot \big(T_\text{a}\left(t, h\right)+273.15\big)},
\end{align}
where $R_\text{d}$ and $R_\text{v}$ are the specific gas constants for dry air and water vapor, respectively. Next, $p_\text{d}$ and $p_\text{v}$ are the pressures of dry air and water vapor. The latter at the altitude $h$ and in the time step $t$ can be expressed by the formula \cite{Omni}:
\begin{align}
    \label{eqVaporPressure}
    p_\text{v}\left(t, h\right)=6.1078\cdot 10^{\frac{7.5\cdot T_\text{a}\left(t, h\right)}{T_\text{a}\left(t, h\right)+237.3}},
\end{align}
The pressure of dry air at the same altitude and moment has been described by $p_\text{d}\left(t, h\right)=p\left(t, h\right)-p_\text{v}\left(t, h\right)$, 
% In turn, $p_\text{d}$ at the specific altitude $h$ and current time step $t$ has been described by the mathematical equation: 
% \begin{align}
%     p_\text{d}\left(t, h\right)=p\left(t, h\right)-p_\text{v}\left(t, h\right),
%     \label{eqDryAirPressure}
% \end{align}
where $p$ is the air pressure evaluated as \cite{Omni}: 
\begin{align}
    \label{eqPressureAtAltitude}
    p\left(t, h\right)=p_{0}\left(t\right)\cdot e^{\frac{-g\left(h\right)\cdot M\cdot \left(h+h_\text{T}-h_{0}\right)}{R\cdot T_\text{a}\left(t, h\right)}},
\end{align}
where $p_{0}$ is the air pressure at the reference level $h_{0}$. It was assumed that the reference level is the sea level altitude $\left(h_{0}=0\right.)$. The parameter of $h_\text{T}$ is the absolute altitude of the terrain. Next referring to \cite{vCalc}, the gravitational acceleration is described by $g\left(h\right)=g_0\frac{r_\text{e}^{2}}{\left(r_\text{e}+h\right)^{2}}$, where $g_0$ and $r_\text{e}$ are the sea level acceleration and mean radius of the Earth, respectively. Finally, the formula to calculate the ambient temperature $\left(T_\text{a}\right)$ at the altitude $h$ and moment $t$ is shown below \cite{Omni}:
\begin{align}
    \label{eqTemperatureAtAltitude}
    T_\text{a}\left(t, h\right)=T_\text{a}(t, h_\text{WS})-0.0065\left(h+h_\text{T}-h_\text{WS}\right),
\end{align}
where $h_\text{WS}$ is the absolute altitude, at which the measurements of weather conditions have been done (the altitude of the weather station -- WS).
\section{Simulation Setup}
\label{section:simulation}
The source code of the developed software was prepared in Java language. The examination of the system scenario described in Section \ref{section:scenario} has been performed in the form of $10$ independent simulation runs each considering $4$ days of the previous year starting different seasons -- vernal equinox $\left(20^\text{th} \text{ March } 2022\right)$, summer solstice $\left(21^\text{st} \text{ June } 2022\right)$, autumn equinox $\left(23^\text{rd} \text{ September } 2022\right)$, and winter solstice $\left(21^\text{st} \text{ December } 2022\right)$. The parameters of users (location coordinates and traffic demand) have always been defined at the beginning of each simulation run. The assumed time step was equal to $1$ minute $\left(4\cdot24\cdot60=5,760 \text{ steps per simulation run}\right)$, with which the weather data was updated, and then the calculations for energy production and consumption (proesumption) were carried out. The simulation setup for network and energy designs is highlighted in Tab.~\ref{table:network_configuration} and \ref{table:prosumption_configuration}. 

\begin{table}[!h]
\centering
\caption{Network Design Configuration \cite{Castellanos, Björnson, PoznanData, NetworkData}}
\label{table:network_configuration}
\resizebox{0.48\textwidth}{!}{
\begin{tabular}{|c|c|c|c|cc|}
\hline
\multirow{2}{*}{}             & \multirow{2}{*}{Parameter} & \multirow{2}{*}{Sign} & \multirow{2}{*}{Unit}     & \multicolumn{2}{c|}{Value}                             \\ \cline{5-6} 
                              &                            &                       &                           & \multicolumn{1}{c|}{BS}                 & UE           \\ \hline
\multirow{4}{*}{\rotatebox[origin=c]{90}{Overall}}      & Quantity                   & $K$                   & --                        & \multicolumn{1}{c|}{$8$}                & $100$        \\ \cline{2-6} 
                              & Movement Speed             & $v$                   & $\left[\text{m/s}\right]$ & \multicolumn{1}{c|}{$\text{N/A}$}       & $0$          \\ \cline{2-6} 
                              & Placement                  & --                    & --                        & \multicolumn{1}{c|}{$\text{N/A}$}       & outdoor      \\ \cline{2-6} 
                              & Technology                 & --                    & --                        & \multicolumn{1}{c|}{$5\text{G}$}        & $\text{N/A}$ \\ \hline
\multirow{11}{*}{\rotatebox[origin=c]{90}{Band}}        & Frequency                  & $f$                   & $\left[\text{MHz}\right]$ & \multicolumn{1}{c|}{$3500$}             & $\text{N/A}$ \\ \cline{2-6} 
                              & Channel Bandwidth          & $B_\text{w}$          & $\left[\text{MHz}\right]$ & \multicolumn{1}{c|}{$120$}              & $\text{N/A}$ \\ \cline{2-6} 
                              & Used Subcarriers           & $N_\text{SC,u}$       & --                        & \multicolumn{1}{c|}{$320$}              & $\text{N/A}$ \\ \cline{2-6} 
                              & Total Subcarriers          & $N_\text{SC,t}$       & --                        & \multicolumn{1}{c|}{$512$}              & $\text{N/A}$ \\ \cline{2-6} 
                              & Sampling Factor            & $\text{SF}$           & --                        & \multicolumn{1}{c|}{$1.536$}            & $\text{N/A}$ \\ \cline{2-6} 
                              & Pilot Reuse Factor         & $\text{RF}$           & --                        & \multicolumn{1}{c|}{$1$}                & $\text{N/A}$ \\ \cline{2-6} 
                              & Coherence Time             & $t_\text{c}$          & $\left[\text{ms}\right]$  & \multicolumn{1}{c|}{$50$}               & $\text{N/A}$ \\ \cline{2-6} 
                              & Coherence Bandwidth        & $B_\text{c}$          & $\left[\text{MHz}\right]$ & \multicolumn{1}{c|}{$1$}                & $\text{N/A}$ \\ \cline{2-6} 
                              & TDD Duty Cycle DL          & $D_\text{DL}$         & $\left[\text{\%}\right]$  & \multicolumn{1}{c|}{$75$}               & $\text{N/A}$ \\ \cline{2-6} 
                              & TDD Duty Cycle UL          & $D_\text{UL}$         & $\left[\text{\%}\right]$  & \multicolumn{1}{c|}{$25$}               & $\text{N/A}$ \\ \cline{2-6} 
                              & Spatial Duty Cycle         & $S$                   & $\left[\text{\%}\right]$  & \multicolumn{1}{c|}{$25$}               & $\text{N/A}$ \\ \hline
\multirow{6}{*}{\rotatebox[origin=c]{90}{Transceivers}} & Antenna Height             & $h$                   & $\left[\text{m}\right]$   & \multicolumn{1}{c|}{$50$}               & $1.5$        \\ \cline{2-6} 
                              & Antenna Elements           & $M$                   & --                        & \multicolumn{1}{c|}{$64$}               & $1$          \\ \cline{2-6} 
                              & Antenna Gain               & $G_\text{a}$          & $\left[\text{dBi}\right]$ & \multicolumn{1}{c|}{$24$}               & $0$          \\ \cline{2-6} 
                              & Antenna Feeder Loss        & $L_\text{f}$          & $\left[\text{dB}\right]$  & \multicolumn{1}{c|}{$3$}                & $0$          \\ \cline{2-6} 
                              & Max. Transmit Power        & $P_\text{TX,max}$     & $\left[\text{dBm}\right]$ & \multicolumn{1}{c|}{$42$}               & $23$         \\ \cline{2-6} 
                              & Noise Figure               & $\text{NF}$           & $\left[\text{dB}\right]$  & \multicolumn{1}{c|}{$7$}                & $\text{N/A}$ \\ \hline
\multirow{7}{*}{\rotatebox[origin=c]{90}{Propagation}}  & Path Loss Model            & --                    & --                        & \multicolumn{1}{c|}{$\text{TR }38.901$} & $\text{N/A}$ \\ \cline{2-6} 
                              & Interference Margin        & $\text{IM}$           & $\left[\text{dB}\right]$  & \multicolumn{1}{c|}{$2$}                & $0$          \\ \cline{2-6} 
                              & Doppler Margin             & $\text{DM}$           & $\left[\text{dB}\right]$  & \multicolumn{1}{c|}{$3$}                & $\text{N/A}$ \\ \cline{2-6} 
                              & Fade Margin                & $\text{FM}$           & $\left[\text{dB}\right]$  & \multicolumn{1}{c|}{$10$}               & $\text{N/A}$ \\ \cline{2-6} 
                              & Shadow Margin              & $\text{SM}$           & $\left[\text{dB}\right]$  & \multicolumn{1}{c|}{$10$}               & $\text{N/A}$ \\ \cline{2-6} 
                              & Implementation Loss        & $\text{IL}$           & $\left[\text{dB}\right]$  & \multicolumn{1}{c|}{$3$}                & $\text{N/A}$ \\ \cline{2-6} 
                              & Soft Handover Gain         & $G_\text{SHO}$        & $\left[\text{dB}\right]$  & \multicolumn{1}{c|}{$\text{N/A}$}       & $0$          \\ \hline
\end{tabular}}
\end{table}

\begin{table}[!t]
\centering
\caption{Energy Prosumption Configuration \cite{Huang, PvData, BatteryData, Björnson, Voltacon, Omni, vCalc, PoznanData, Arnold, Franklin, HomerGloss}}
\label{table:prosumption_configuration}
\resizebox{0.48\textwidth}{!}{
\begin{tabular}{|c|c|c|c|c|}
\hline
                    & Parameter                                                    & Sign                      & Unit       & Value                           \\ \hline
\multirow{7}{*}{\rotatebox[origin=c]{90}{UAV Device}}      & Mass of UAV                           & $m_\text{UAV}$                  & {$\left[\text{kg}\right]$}    & $2$                              \\ \cline{2-5}
                                 & Auxiliary Mass                                               & $m_\text{AUX}$                  & {$\left[\text{kg}\right]$}    & $0$                                \\ \cline{2-5}
                                 & Auxiliary Power                                              & $P_\text{AUX}$                    & {$\left[\text{W}\right]$}    & $0$                               \\ \cline{2-5}
                                 & Hovering Altitude                                      & $h_\text{UAV}$                     & {$\left[\text{m}\right]$}    & $50$                               \\ \cline{2-5}
                                 & Single Propeller Radius                                     & $r_\text{p}$                     & {$\left[\text{m}\right]$}          & $0.5$                              \\ \cline{2-5}
                                 & Number of Propellers                                     & $l_\text{p}$                     & --          & $12$                              \\ \cline{2-5} 
                                 & DC Loss Factor                                               & $\sigma_\text{DC}$                 & --          & $0.075$                           \\ \hline
\multirow{10}{*}{\rotatebox[origin=c]{90}{MIMO Transceiver}} 
                                 & Mass of MIMO Transceiver                                     & $m_\text{MIMO}$                  & {$\left[\text{kg}\right]$}    & $1$                               \\ \cline{2-5}
                                 & Fixed Power Component                                        & $P_\text{FIX}$                    & {$\left[\text{W}\right]$}    & $10$                              \\ \cline{2-5} 
                                 & Local Oscillator Power                                       & $P_\text{LO}$                     & {$\left[\text{W}\right]$}    & $0.2$                             \\ \cline{2-5} 
                                 & Circuit Components Power                                     & $P_\text{CC}$                     & {$\left[\text{W}\right]$}    & $0.4$                             \\ \cline{2-5} 
                                 & Encoding Power                                               & $P_\text{COD}$                    & {$\left[\text{W}\right]$}    & $0.1$                             \\ \cline{2-5} 
                                 & Decoding Power                                               & $P_\text{DEC}$                    & {$\left[\text{W}\right]$}    & $0.8$                             \\ \cline{2-5} 
                                 & Backhaul Traffic Power                                       & $P_\text{BT}$                     & {$\left[\text{W}\right]$}    & $0.25$                            \\ \cline{2-5}  
                                 & Computational Efficiency                                     & $\eta_\text{BS}$                     & {$\left[\text{Gflops}/\text{W}\right]$}          & $75$                              \\ \cline{2-5}
                                 & Amplifier Efficiency                                         & $\mu_\text{PA}$                     & --          & $0.35$ \\ \cline{2-5}
                                 & Number of Transceivers                                     & $N_\text{MIMO}$                     & --          & $1$ \\ \hline
\multirow{5}{*}{\rotatebox[origin=c]{90}{RIS Array}}      & Mass of RIS                                            & $m_\text{RIS}$                  & {$\left[\text{kg}\right]$}    & $1$                                \\ \cline{2-5}
                                 & Phase Shifter Power                                          & $P_\text{PSH}$                    & {$\left[\text{W}\right]$}    & $7.8$                               \\ \cline{2-5}
                                 & Phase Shifter Bit Resolution                                     & $b_\text{PSH}$                     & {$\left[\text{bits}\right]$}          & $6$                              \\ \cline{2-5}
                                 & Number of Reflecting Elements                                     & $N_\text{RE}$                     & --          & $16$ \\ \cline{2-5}
                                 & Number of RISs                                     & $N_\text{RIS}$                     & --          & $1$ \\ \hline
\multirow{19}{*}{\rotatebox[origin=c]{90}{PV Panel}}      & Model                                              & \multicolumn{3}{c|} {Solarland SLP$020$-$12$U}            \\ \cline{2-5}  
                                 & Mass of PV                          & $m_\text{PV}$                  & {$\left[\text{kg}\right]$}    & $0$                              \\ \cline{2-5}
                                 & Nominal Voltage                           & $V_\text{n,PV}$                  & {$\left[\text{V}\right]$}    & $12$                              \\ \cline{2-5} 
                                 & Voltage at Max. Power                       & $V_\text{max,PV}$                & {$\left[\text{V}\right]$}    & $17.2$                            \\ \cline{2-5} 
                                 & Current at Max. Power                       & $I_\text{max, PV}$                & {$\left[\text{A}\right]$}    & $1.16$                            \\ \cline{2-5} 
                                 & Rated Power                               & $P_\text{R,PV}$                  & {$\left[\text{W}\right]$}    & $20$                              \\ \cline{2-5} 
                                 & Ground-relative Altitude                                      & $h_\text{PV}$                     & {$\left[\text{m}\right]$}    & $50$                               \\ \cline{2-5}
                                 & Module Dimensions                 & $a_\text{PV}$ x $b_\text{PV}$             & {$\left[\text{m}\right]$}    & $0.576$ x $0.357$                   \\ \cline{2-5} 
                                 & Solar Radiation at STC                                       & $\overline{G}_\text{T,STC}$                    & {$\left[\text{W/}\text{m}^{2}\right]$}  & $1000$                            \\ \cline{2-5} 
                                 & Solar Radiation for NOCT                                      & $\overline{G}_\text{T,NOCT}$                   & {$\left[\text{W/}\text{m}^{2}\right]$}  & $800$                             \\ \cline{2-5}
                                 & Temperature under STC                                      & $T_\text{c,STC}$                    & {$\left[^\circ\text{C}\right]$}    & $25$                              \\ \cline{2-5} 
                                 & Temperature NOCT                                     & $T_\text{c,NOCT}$                   & {$\left[^\circ\text{C}\right]$}    & $47$                              \\ \cline{2-5} 
                                 & Temperature Coeffi. of Power                             & $\alpha_\text{PV}$                 & {$\left[\text{\%/}^\circ\text{C}\right]$} & $-0.5$                            \\ \cline{2-5} 
                                 & Solar Absorptance                                        & $\alpha$                    & --          & $0.3\sqrt{10}$                            \\ \cline{2-5}
                                 & Solar Transmittance                                          & $\tau$                    & --          & $0.3\sqrt{10}$                            \\ \cline{2-5}
                                 & Derating Factor                                          & $f_\text{PV}$                    & --          & $0.723$                            \\ \cline{2-5} 
                                 & Number in Serial Order                               & $N_\text{PV,s}$                  & --          & $1$                               \\ \cline{2-5} 
                                 & Number in Parallel Order                             & $N_\text{PV,p}$                  & --          & $5$                              \\ \cline{2-5} 
                                 & Total Number per Net. Cell                          & $N_\text{PV}$                     & --          & $5$                              \\ \hline
\multirow{16}{*}{\rotatebox[origin=c]{90}{Battery System}} & Model                                                & \multicolumn{3}{c|} {Volt Accumulator $\text{LiFePO}_{4}$ $12.8$V $60$Ah} \\ \cline{2-5} 
                                 & Mass of Battery                         & $m_\text{BATT}$                  & {$\left[\text{kg}\right]$}    & $5.2$                              \\ \cline{2-5}
                                 & Nominal Voltage                          & $V_\text{n,BATT}$                & {$\left[\text{V}\right]$}    & $12.8$                            \\ \cline{2-5} 
                                 & Charging Voltage                         & $V_\text{c,BATT}$                & {$\left[\text{V}\right]$}    & $14.6$                            \\ \cline{2-5} 
                                 & Discharging Voltage                      & $V_\text{d,BATT}$                & {$\left[\text{V}\right]$}    & $12.8$                            \\ \cline{2-5} 
                                 & Charging Current                         & $I_\text{c,BATT}$                & {$\left[\text{A}\right]$}    & $30$                              \\ \cline{2-5} 
                                 & Discharging Current                      & $I_\text{d,BATT}$                & {$\left[\text{A}\right]$}    & $60$                              \\ \cline{2-5} 
                                 & Capacity                                 & $C_\text{BATT}$                   & {$\left[\text{Ah}\right]$}   & $60$                           \\ \cline{2-5} 
                                 & Provided Energy                          & $E'_\text{BATT,max}$                   & {$\left[\text{Wh}\right]$}   & $768$                            \\ \cline{2-5} 
                                 & Max. Depth of Discharge                  & $\text{DoD}_\text{max}$                  & {$\left[\text{\%}\right]$}   & $100$                             \\ \cline{2-5} 
                                 & Primary State of Charge                  & $\text{SoC}_\text{p}$                    & {$\left[\text{\%}\right]$}   & $95$                             \\ \cline{2-5} 
                                 & Battery's Efficiency                                 & $\mu_\text{BATT}$                   & --          & $0.95$                        \\ \cline{2-5}    
                                 & Number of Cycles                         & $N_\text{BC}$                     & --          & $2000$                            \\ \cline{2-5} 
                                 & Number in Serial Order                                & $N_\text{BATT,s}$                 & --          & $1$                               \\ \cline{2-5} 
                                 & Number in Parallel Order                             & $N_\text{BATT,p}$                 & --          & $1$                               \\ \cline{2-5}
                                 & Total Number per Net. Cell                          & $N_\text{BATT}$                   & --          & $1$                               \\ \hline 
\multirow{9}{*}{\rotatebox[origin=c]{90}{Other}} & Reference Altitude                                        & $h_{0}$                    & {$\left[\text{m}\right]$}    & $0$                              \\ \cline{2-5} 
                                 & Terrain Absolute Altitude                                       & $h_\text{T}$                     & {$\left[\text{m}\right]$}    & $54.44$                             \\ \cline{2-5} 
                                 & Weather Station Absolute Alti.                                     & $h_\text{WS}$                     & {$\left[\text{m}\right]$}    & $90$                             \\ \cline{2-5}
                                 & Mean Radius of the Earth                                     & $r_\text{e}$                     & {$\left[\text{m}\right]$}    & $6371009$                             \\ \cline{2-5}
                                 & Sea Level Gravitational Accel.                                     & $g_{0}$                     & {$\left[\text{m}/\text{s}^{2}\right]$}    & $9.80665$                             \\ \cline{2-5}
                                 & Air Molar Mass                                               & $m_\text{air}$                    & {$\left[\text{kg}/\text{mol}\right]$}    & $0.0289644$                             \\ \cline{2-5}
                                 & Universal Gas Constant                                     & $R_\text{u}$                     & {$\left[\frac{\text{N}\cdot \text{m}}{\text{mol}\cdot \text{K}}\right]$}    & $8.31432$                             \\ \cline{2-5} 
                                 & Dry Air Gas Constant                                     & $R_\text{d}$                     & {$\left[\text{J}/\left(\text{kg}\cdot\text{K}\right)\right]$}    & $287.058$                             \\ \cline{2-5} 
                                 & Water Vapor Gas Constant                                     & $R_\text{v}$                     & {$\left[\text{J}/\left(\text{kg}\cdot\text{K}\right)\right]$}    & $461.495$                             \\ \hline
\end{tabular}}
\end{table}

\section{Results}
\label{section:results}
The results of performed simulations have been attached within Tab.~\ref{table:energy_characteristics}. The first array presents the amount of energy that can be harvested by PV panels of a single UAV during the whole year on average detailing each season. According to the initial expectations, the biggest amount of resources the mobile base station is able to obtain from solar radiation is the summer solstice $\left(572.64\right)$, where the peak value of the energy production process is also the highest $\left(91.86\right)$. The ranking was followed by vernal and autumn equinoxes and winter solstice. Hence, there could be seen that in terms of the reduction of energy delivered by the conventional sources (i.e., from the batteries, which are charged up from the dedicated stations) the order is adequate to the aforementioned dependencies (middle array). However, due to the limitations related to the number of PV cells as well as their efficiency of power generation, the maximal achieved energy gain was equal to $5.8\%$ (summer solstice). It is also valid to be noticed that during the winter solstice, this gain is almost none $(0.18\%)$. Finally, the bottom array indicates the average number of UAV BS replacements, when its battery is gone. Due to weather conditions, the variety of this number can even be observed when are no RESs engaged to power up the mobile access node. The highest number of exchanged drones very noticed for summer solstice $\left(13.93\right)$ and next for autumn equinox $\left(13.39\right)$, winter solstice $\left(13.28\right)$, and vernal equinox $\left(13.24\right)$, respectively. On the contrary, when the UAVs are supported by PV panels, summer solstice as well as vernal equinox needs the lowest average number of replacements $\left(12.98\right)$. Taking into account the fact that during the summer the energy demand of a single UAV BS increases compared to other seasons of the year, this confirms the above-described results related to the profit when generating resources from solar radiation at this time. Thus, due to the almost zero impact of using PV arrays in winter solstice on power consumption characteristics, the number of UAV replacements is the same for both cases, i.e., with and without enabled RESs.
\begin{table}[!h]
\centering
\caption{Energy characteristics for UAV BSs with and without RESs}
\label{table:energy_characteristics}
\resizebox{0.48\textwidth}{!}{
\begin{tabular}{|c|cc|}
\hline
                                      & \multicolumn{2}{c|}{Total (and peak) energy obtained from PV Panels per UAV $\left[\text{Wh}\right]$} \\ \cline{2-3}
\multirow{-2}{*}{}                    & \multicolumn{1}{c|}{\quad\quad\quad\textit{No RESs}\quad\quad\quad\quad}       & \multicolumn{1}{c|}{\textit{PV Panels}}         \\ \hline
Vernal Equinox                        & \multicolumn{1}{c|}{$0 \text{ } \left(0\right)$}        & $475.17 \text{ } \left(60.57\right)$         \\ \hline
Summer Solstice                       & \multicolumn{1}{c|}{$0 \text{ } \left(0\right)$}        & $572.64 \text{ } \left(91.86\right)$         \\ \hline
Autumn Equinox                        & \multicolumn{1}{c|}{$0 \text{ } \left(0\right)$}        & $349.56 \text{ } \left(65.15\right)$         \\ \hline
Winter Solstice                       & \multicolumn{1}{c|}{$0 \text{ } \left(0\right)$}        & $17.67 \text{ } \left(4.18\right)$         \\ \hline
\rowcolor[HTML]{6665CD} 
{\color[HTML]{FFFFFF} Annual average} & \multicolumn{1}{c|}{\cellcolor[HTML]{6665CD}{\color[HTML]{FFFFFF} $0$}} & {\color[HTML]{FFFFFF} $353.76$} \\ \hline
\end{tabular}}

\begin{tabular}{cc}
     \\
\end{tabular}

\resizebox{0.48\textwidth}{!}{
\begin{tabular}{|c|cc|}
\hline
                                      & \multicolumn{2}{c|}{Average reduction in energy consumption (AREC) $\left[\text{\%}\right]$} \\ \cline{2-3}
\multirow{-2}{*}{}                    & \multicolumn{1}{c|}{\quad\quad\quad\textit{No RESs}\quad\quad\quad\quad}       & \multicolumn{1}{c|}{\textit{PV Panels}}         \\ \hline
Vernal Equinox                        & \multicolumn{1}{c|}{$0$}                                                & $4.89$                        \\ \hline
Summer Solstice                       & \multicolumn{1}{c|}{$0$}                                                & $5.8$                        \\ \hline
Autumn Equinox                        & \multicolumn{1}{c|}{$0$}                                                & $3.56$                        \\ \hline
Winter Solstice                       & \multicolumn{1}{c|}{$0$}                                                & $0.18$                        \\ \hline
\rowcolor[HTML]{6665CD} 
{\color[HTML]{FFFFFF} Annual average} & \multicolumn{1}{c|}{\cellcolor[HTML]{6665CD}{\color[HTML]{FFFFFF} $0$}} & {\color[HTML]{FFFFFF} $3.61$} \\ \hline
\end{tabular}}

\begin{tabular}{cc}
     \\
\end{tabular}

\resizebox{0.48\textwidth}{!}{
\begin{tabular}{|c|cc|}
\hline
                                      & \multicolumn{2}{c|}{Avarege number of UAV replacements (ANUR)} \\ \cline{2-3}
\multirow{-2}{*}{}                    & \multicolumn{1}{c|}{\quad\quad\quad\textit{No RESs}\quad\quad\quad\quad}       & \multicolumn{1}{c|}{\textit{PV Panels}}         \\ \hline
Vernal Equinox                        & \multicolumn{1}{c|}{$13.24$}                                                & $12.98$                        \\ \hline
Summer Solstice                       & \multicolumn{1}{c|}{$13.93$}                                                & $12.98$                        \\ \hline
Autumn Equinox                        & \multicolumn{1}{c|}{$13.39$}                                                & $13.09$                        \\ \hline
Winter Solstice                       & \multicolumn{1}{c|}{$13.28$}                                                & $13.28$                        \\ \hline
\rowcolor[HTML]{6665CD} 
{\color[HTML]{FFFFFF} Annual average} & \multicolumn{1}{c|}{\cellcolor[HTML]{6665CD}{\color[HTML]{FFFFFF} $13.46$}} & {\color[HTML]{FFFFFF} $13.08$} \\ \hline
\end{tabular}}
\end{table}

\section{Conclusions}
\label{section:conclusions}
The contribution presented in this paper highlights the advantages related to the use of PV panels as power generators in cellular networks equipped with UAVs as mobile access nodes and supported by RISs. For the considered scenario, due to the weather conditions prevailing in Poland as well as the assumed configurations of UAVs and RESs, the power savings (and the resulting financial ones) are equal to the level of $3.61\%$ per year on average in comparison to the case, in which base stations of the wireless system are supplied only from the charging stations powered by the conventional energy grid. Although RESs like PV panels are characterized by time-varying and climate-dependent harvesting processes, by appropriate management of available resources (radio and energy) using optimizing algorithms (e.g., traffic steering, resource allocation, etc.) and enabling additional equipment like RIS arrays, we are able to improve already achieved results or even ensure energy autonomy for cellular network without worsening the quality of mobile services delivered to users. However, the implementation of those algorithms as well as studies focused on the impact of RIS on radio signal propagation will be taken into consideration in future work.

\section*{Acknowledgment}
\label{section:acknowledgment}

The authors would like to thank Prof. Margot Deruyck from the Ghent University -- IMEC in Belgium for supporting the following work by providing the GRAND tool. The work was realized within project no. $2021/43/\text{B}/\text{ST}7/01365$ funded by National Science Center in Poland.

% \section*{References}
% \label{section:references}

\vspace{12pt}

\end{document}